\documentstyle[aps,preprint,epsfig,floats]{revtex}

\begin{document}

\hfuzz=4pt

\preprint{\font\fortssbx=cmssbx10 scaled \magstep2
\hbox to \hsize{
\hbox{\fortssbx University of Wisconsin - Madison}
\hfill$\vcenter{\tighten
                \hbox{\bf MADPH-97-995}
                \hbox{\bf IFT-P.039/97}
                \hbox{\bf IFUSP-1270/97}
                \hbox{\bf hep-ph/9708283}
                \hbox{August 1997}}$}}

\title{\vspace*{.15in}
Are Two Gluons the QCD Pomeron?}

\author{O.\ J.\ P.\ \'Eboli}
\address{Instituto de F\'{\i}sica, Universidade de S\~ao Paulo \\
C.P.\ 66.318, 05315--970, S\~ao Paulo -- SP, Brazil}

\author{E.\ M.\ Gregores}
\address{Instituto de F\'{\i}sica Te\'orica,
Universidade Estadual Paulista \\
Rua Pamplona 145, 01405--900, S\~ao Paulo -- SP, Brazil}

\author{F.\ Halzen}
\address{Department of Physics, University of Wisconsin \\
Madison, WI 53706, USA}

\maketitle
\thispagestyle{empty}

\begin{abstract}
\vskip-5ex 

Collider experiments have recently measured the production cross
section of hard jets separated by a rapidity gap as a function of
transverse momentum and gap size.  We show that these measurements
reveal the relative frequencies for the production of rapidity gaps in
quark-quark, quark-gluon and gluon-gluon interactions. The results are
at variance with the idea that the exchange of 2 gluons is a first
order approximation for the mechanism producing colorless states, i.e.\
the hard QCD Pomeron. They do qualitatively support the ``soft color"
or ``color evaporation" scheme developed in the context of bound-state
heavy quark production.

\end{abstract}


\section{Introduction:}

Although we do understand strong interactions in the context of QCD,
we can at best speculate on how to calculate elastic scattering. This
is just one example of an interaction mediated by the exchange of the
``Pomeron", a state which carries no net color. The routine
speculation has been that it is, to a first approximation, a state of
two colored gluons combined into a color singlet. Understanding the
Pomeron has been challenging because its dynamics is revealed in
processes which are not subject to perturbative computation, e.g.
elastic scattering. The hope has been that it may be instructive to
study Pomeron dynamics in hard processes that are, at least partially,
understood in terms of perturbative QCD: the hard Pomeron. Examples
include processes involving colorless pairs of heavy quarks, e.g.
$\psi$'s, or colorless states of light quarks produced in association
with a pair of high transverse momentum jets: rapidity gaps. In this
paper we will show how a treatment of colorless states in QCD,
suggested by the phenomenology of heavy quark bound states, supports a
``soft color" model of the Pomeron which is at variance with the idea
that it is a structure built on a frame composed of two gluons. We
will show that recent measurements\cite{jill} of the relative frequency
for the production of rapidity gaps in quark-quark, quark-gluon and
gluon-gluon interactions provides qualitative, yet convincing,
confirmation of the soft color concept.

The reason why some data on the production of $\psi$- and
$\Upsilon$-states radically disagree with QCD predictions,
occasionally by well over one order of magnitude, is that the
traditional method for performing the perturbative calculation of the
cross section is simply wrong\cite{faro}. The key mistake is to
require that the heavy quark pair forms a color singlet at short
distances, given that there is an essentially infinite time for soft
gluons to readjust the color of the $c \bar c$ pair before it appears
as an asymptotic $\psi$ or, alternatively, $D \bar D$ state. We
suspect that the same mistake is made in the description of rapidity
gaps, i.e.\ the production of a color-neutral quark-antiquark pair, in
terms of the exchange of a color neutral gluon pair. The $\psi$ is,
after all, a color neutral $c \bar c$ pair and we have
shown\cite{faro} in quantitative detail that it is indeed produced by
exactly the same dynamics as $D \bar D$ pairs; its color happens to be
bleached by soft final-state interactions. This approach to color is
also suggestive of the unorthodox prescription for the production of
rapidity gaps in deep inelastic scattering, proposed by Buchm\"uller
and Hebecker\cite{bh}.

In this paper we emphasize that recent measurements\cite{jill} on the
formation of rapidity gaps between a pair of high transverse momentum
jets shed new light on the problem of how to treat color in semi-hard
interactions. When applied to the formation of gaps between a pair of
high transverse momentum jets in hadron collisions, the ``soft color"
or ``color evaporation" approach suggests a formation rate of gaps in
gluon-gluon subprocesses which is similar, or smaller, than in
quark-quark induced events. Consequently, formation of gaps should
increase with increased transverse momentum, or reduced collision
energy of the jets. This prediction happens to be antithetical to the
one obtained in 2-gluon exchange Pomeron models. We show that the data
resolve the issue in favor of the ``soft color" computational scheme
and questions the relevance of approximating the exchange of a
color-singlet, hard Pomeron as a pair of gluons. We also exhibit the
predictions of the ``soft color'' model for the production of rapidity
gaps at the LHC energy.


\section{Onium Calculations with Soft Color: a Brief Reminder}

The conventional treatment of color in perturbative QCD calculations,
i.e., the color singlet model, has run into serious problems in
describing the data on the production of charmonium and upsilon
states\cite{review}. Specific proposals to solve the onium problem
agree on the basic solution: onium production is a two-step process
where a heavy quark pair is produced first. At this initial stage all
perturbative diagrams are included, whether the $c \bar c$ pair forms
a color singlet state or not. This is a departure of the textbook
approach where only diagrams with the charm pair in a color singlet
state are selected. In the Bodwin-Braaten-Lepage (BBL)
formalism\cite{bbl} the subsequent evolution of the pair into a
colorless bound state is described by an expansion in powers of the
relative velocity of the heavy quarks in the onium system. A different
approach, the color evaporation or soft color method, represents an
even more radical departure from the way color singlet states are
conventionally treated in perturbation theory.  Color is, in fact,
``ignored''.  Rather than explicitly imposing that the system is in a
color singlet state in the short-distance perturbative diagrams, the
appearance of color singlet asymptotic states depends solely on the
outcome of large-distance fluctuations of quarks and gluons. In other
words, color is controlled by nonperturbative interactions.

\begin{figure}
\parbox[c]{3.0in}{
\mbox{\qquad\epsfig{file=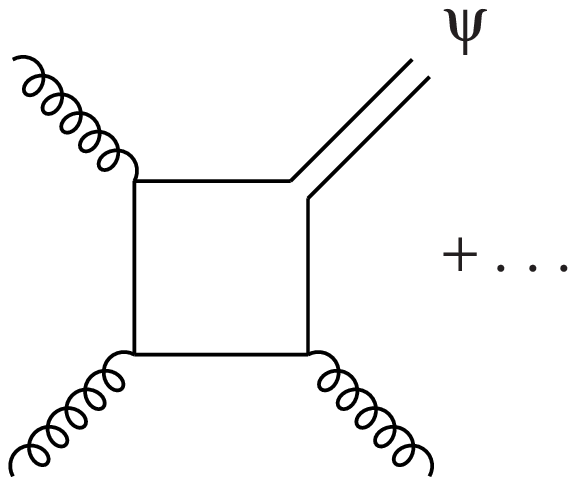,width=.75\linewidth}}
\begin{center}{\bf (a)~~~~~~~~}\end{center}
  }
\hfill
\parbox[c]{3.0in}{
\mbox{\epsfig{file=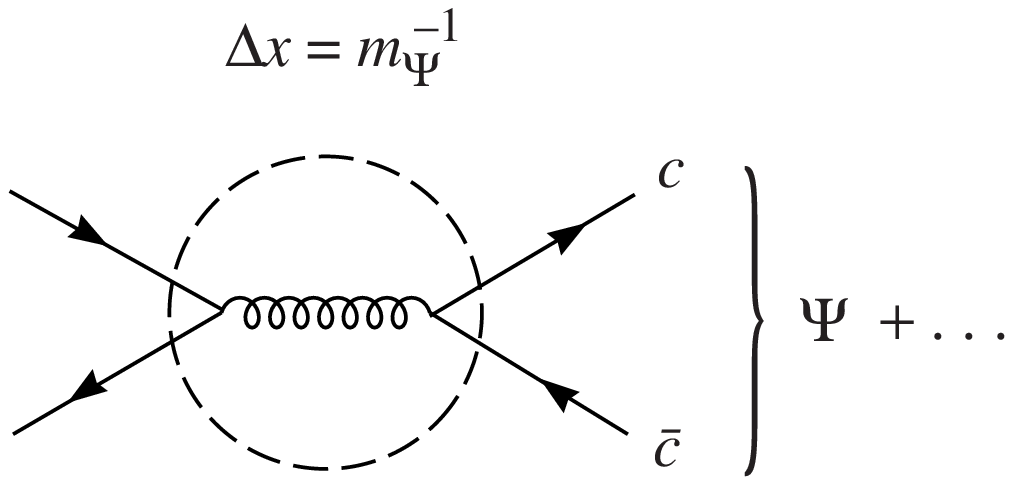,width=\linewidth}}
\begin{center}{\bf (b)~~~~~~~~~~~}\end{center}
}
\caption{Quarkonium production: (a) Color Singlet Model; (b) Color
Evaporation Model.}
\label{fig:csm}
\end{figure}

In Fig.~\ref{fig:csm} we show typical diagrams for the production of
$\psi$-particles representing the competing treatments of the color
quantum number. In the diagram of Fig.~\ref{fig:csm}a, the color
singlet approach, the $\psi$ is produced in gluon-gluon interactions
in association with a final state gluon, which is required by color
conservation. This diagram is related by crossing to the hadronic
decay $\psi \rightarrow 3$ gluons. In the color evaporation approach,
the color singlet property of the $\psi$ is ignored at the
perturbative stage of the calculation. The $\psi$ can, for instance,
be produced to leading order by $q\bar q$-annihilation into $c\bar c$,
which is the color-equivalent of the Drell-Yan process, as shown in
Fig.~\ref{fig:csm}b.  This diagram is calculated perturbatively, with
dynamics dictated by short-distance interactions of range $\Delta x
\simeq m_{\psi}^{-1}$.  It does indeed not seem logical to enforce the
color singlet property of the $\psi$ at short distances, given that
there is an essentially infinite time for soft gluons to readjust the
color of the $c \bar c$ pair before it appears as an asymptotic $\psi$
or, alternatively, $D \bar D$ state. Alternatively, it is indeed hard
to imagine that a color singlet state formed at a range
$m_{\psi}^{-1}$, automatically survives to form a $\psi$. This
formalism represents the original\cite{cem,fh:1a,fh:1b,gor} and, as we
have shown\cite{amundson}, correct method by which perturbative QCD
calculations should be performed.

The evidence is compelling that Nature operates according to the color
evaporation scheme. The formalism predicts that, up to color and
normalization factors, the energy, $x_F$- and $p_T$-dependences of the
cross section, are identical for the production of onium states and $D
\bar D$ pairs. This is indeed the case\cite{amundson,oscar}.  Another
striking feature is that the production of charmonium is dominated by
the conversion of a colored gluon into a $\psi$, as in
Fig.~\ref{fig:csm}b.  In the conventional treatment, where the color
singlet property of the $\psi$ is enforced at the perturbative level,
3 gluons (or 2 gluons and a photon) are required to produce a $\psi$.
Contrary to the usual folklore, $\psi$'s are not produced only by
gluon--gluon interactions. As a consequence color evaporation predicts
an enhanced $\psi$ cross section for antiproton beams, while the color
singlet model predicts roughly equal cross sections for proton and
antiproton beams. The prediction of an enhanced $\bar p$ yield is
obviously correct: antiproton production of $\psi$'s exceeds that by
protons by a factor 5 close to threshold. This fact has been known for
some time\cite{fh:1a,fh:1b,gor}. We should note that for sufficiently
high energies, gluon initial states will eventually dominate because
they represent the bulk of soft partons.

Quantitative tests of color evaporation are made possible by the fact
that all $\psi$-production data, i.e.\ photo-, hadroproduction,
$Z$-decay, etc., are described in terms of a single parameter: the
parameter determining the frequency by which a charm pair turns into a
$\psi$ via the final state color fluctuations. Once this parameter has
been empirically determined for one initial state, the cross section
is predicted without free parameters for any other.  We have
demonstrated\cite{oscar} the quantitative precision of the color
evaporation scheme by showing how it accommodates all measurements,
including the high energy Tevatron and HERA data, which have
represented a considerable challenge for other computational schemes. Its
parameter-free prediction of the rate for $Z$-boson decay into $\psi$'s
is an order of magnitude larger than the color singlet model and
consistent with data\cite{oscarZ}.

In summary, the soft color approach gives a complete picture of
charmonium production in hadron-hadron, $\gamma$-hadron, and $Z$
decays. The phenomenological success of the soft color scheme is
impressive and extends to applications to other charmonium and upsilon
states\cite{amundson,schuler}.


\section{Rapidity Gaps as Colorless States of (Light) Quarks and Gluons}

\begin{figure}
\parbox[c]{3.0in}{
\mbox{\epsfig{file=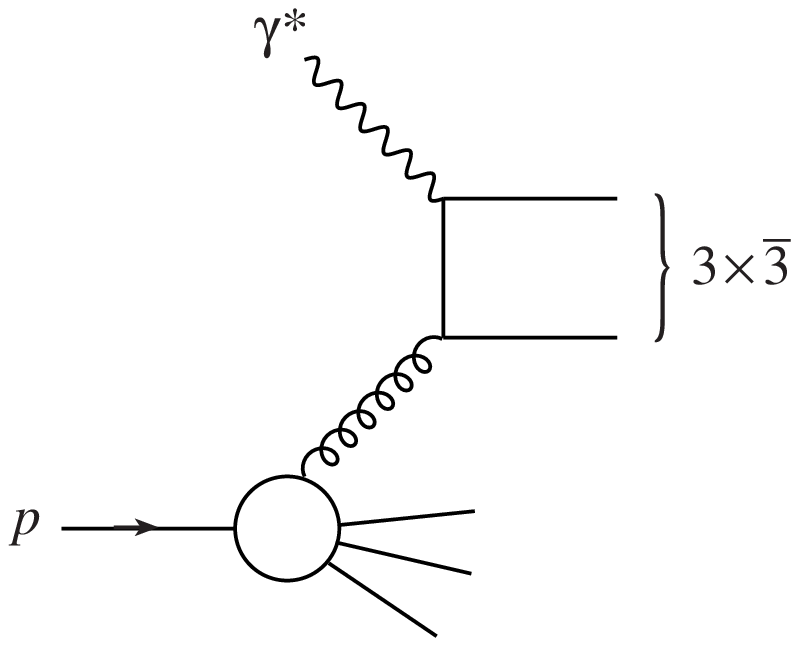,width=\linewidth}}
\begin{center}{\bf (a)}\end{center}
  }
\hfill
\parbox[c]{3.0in}{
\mbox{\epsfig{file=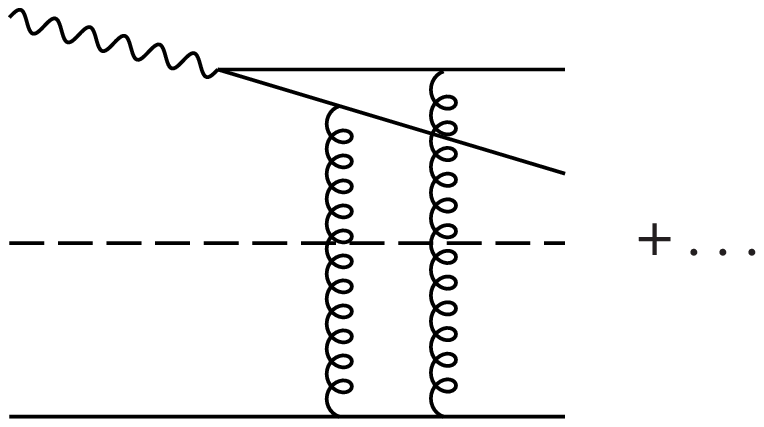,width=\linewidth}}
\begin{center}{\bf (b)~~~~~}\end{center}
}
\caption{Rapidity gaps in deep inelastic scattering: (a) due to soft colors
effects; (b) due to pomeron exchange.}
\label{gap:dis}
\end{figure}

We now turn to the implications of the soft color scheme for the
dynamics underlying the production of rapidity gaps, which refer to
regions in phase space where no hadrons appear as a result of the
production of a color neutral partonic system. The connection to
charmonium physics is obvious: the $\psi$ is a color-neutral $c
\bar{c}$ pair.  The important lesson from heavy quark phenomenology is
that perturbative color octet states fully contribute to the
asymptotic production of color singlet states, such as $\psi$'s. We
suspect that this is also true for the production of a rapidity gap
which represents nothing but the creation of a color singlet
quark-antiquark pair, as shown in Fig.~\ref{gap:dis}a for
electroproduction.  The diagram represents the production of final
state hadrons which are ordered in rapidity: From top to bottom we
find the fragments of the intermediate partonic quark-antiquark state
and those of the target.  Buchm\"uller and Hebecker proposed that the
origin of a rapidity gap corresponds to the absence of a color string
between photon and proton remnants, {\em i.e.} the $\mbox{\bf 3}
\times \bar{\mbox{\bf 3}}$ ($= \mbox{\bf 1} + \mbox{\bf 8}$)
intermediate quark-antiquark state is in a color singlet state.
Because color is the source of hadrons, only the color octet states
yield hadronic asymptotic states.  The reasonable guess that
\begin{equation}
        F_2^{(gap)} = \frac{1}{1+8} F_2
\end{equation}
follows from this argument.

Although this is only a guess, it embodies the essential physics:
events with and without gaps are described by the same short-distance
dynamics. Essentially non-perturbative final-state interactions
dictate the appearance of gaps whose frequency may, possibly, be
determined by simple counting. The treatment of color is the same as
in the case of heavy quark production and leads to similar
predictions: the same perturbative mechanisms, i.e.\ gluon exchange,
dictates the dynamics of color-singlet gap ($\psi$) and regular deep
inelastic (open charm) events.

\begin{figure}
\parbox[c]{3.0in}{
\mbox{\epsfig{file=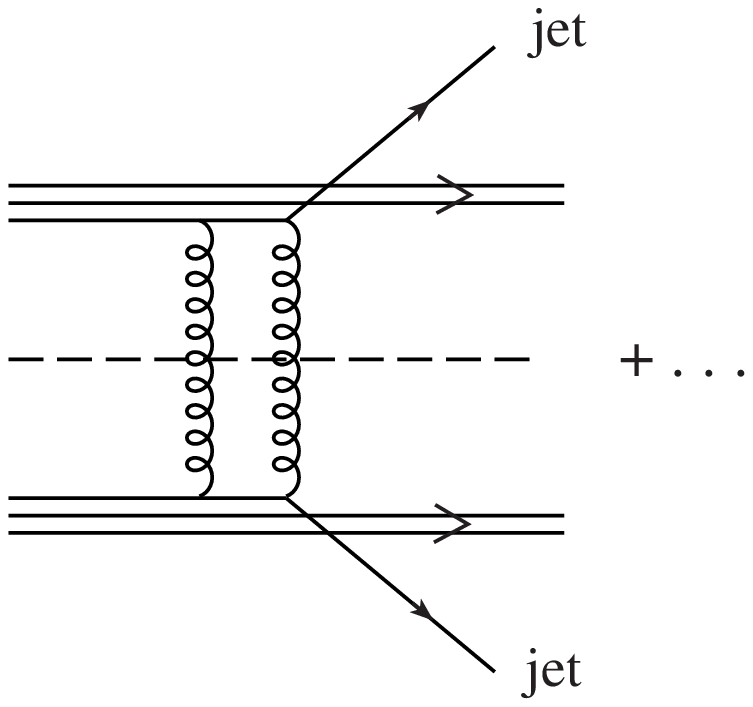,width=\linewidth}}
\begin{center}{\bf (a)}\end{center}
  }
\hfill
\parbox[c]{3.0in}{
\mbox{\epsfig{file=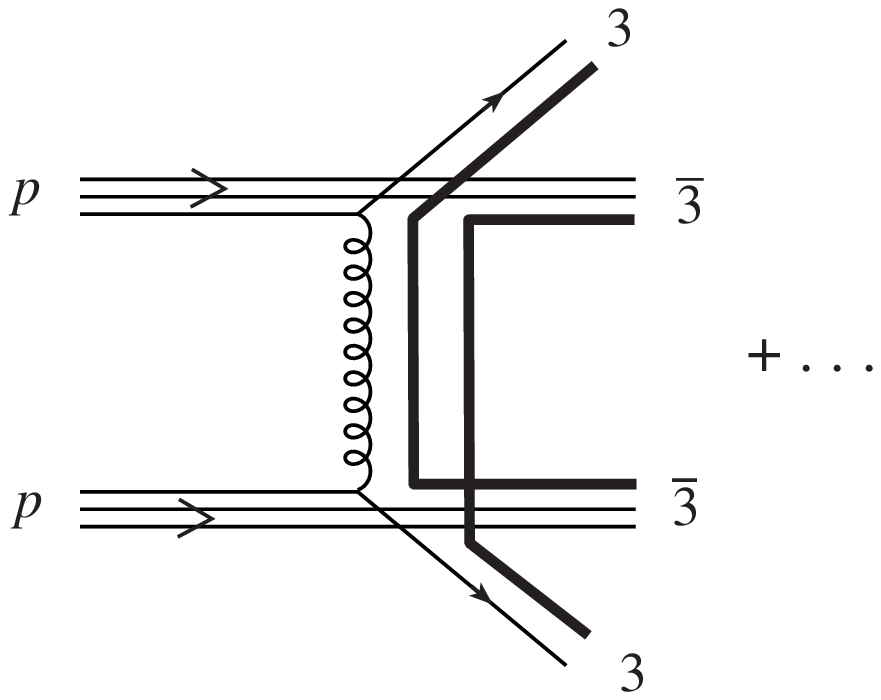,width=\linewidth}}
\begin{center}{\bf (b)}\end{center}
}
\caption{Rapidity gap between jets: (a) due to Pomeron exchange;
(b) due to soft color interactions.}
\label{2j:pom}
\end{figure}

Our proposal for the (soft) nature of color challenges the orthodox
mechanism for producing rapidity gaps sketched in Fig.~\ref{gap:dis}b,
where the $t$-channel exchange of a pair of gluons in a color singlet
state is the origin of the gap. The color string which connects photon
and proton remnants in diagrams such as the one in
Fig.~\ref{gap:dis}a, is absent and no hadrons are produced in the
rapidity region separating them.  The same mechanism predicts rapidity
gaps between a pair of jets produced in hadronic collisions; see
Fig.~\ref{2j:pom}a. These have been observed and occur with a
frequency of order of one percent\cite{prl1,cdf1,prl2}.

The arguments developed in this paper question the hard Pomeron
approach: it is as meaningless to enforce the color singlet nature of
the gluon pair as it is to require that the $c \bar c$ pair producing
a $\psi$ is colorless at the perturbative level. Following our color
scheme the gaps originate from a mere final state color bleaching
phenomenon {\em\`{a} la} Buchm\"uller and Hebecker. This can be
visualized using the diagram shown in Fig.~\ref{2j:pom}b.  At short
distances it represents a conventional perturbative diagram for the
production of a pair of jets. Therefore, the same short distance
dynamics governs events with and without rapidity gaps, as was the
case for electroproduction. This is consistent with all experimental
information.

Producing a quantitative model for the gap rate may be premature at
this point. There are at least two consistent interpretations of the
present data. The first is based on the string picture for the
formation of the final state hadrons shown in Fig.~\ref{2j:pom}b.
Color in the final state is bleached by strings connecting the
${\mbox{\bf 3}}$ jet at the top with the $\bar{\mbox{\bf 3}}$
spectator di-quark at the bottom and vice-versa, resulting in color
singlet states at the top and bottom.  The probability to form a gap
can be counted {\em \`{a} la} Buchm\"uller and Hebecker to be
$1/(1+8)^2$ because it requires the formation of singlets in 2
strings.  The data\cite{prl1,cdf1,prl2} is consistent with this simple
picture which basically predicts that the gap fraction between
$p\bar{p}$ jets is the square of that between virtual photon and
proton in deep inelastic scattering. One could, alternatively, argue
that, once color has been bleached between one
${\mbox{\bf3}}$--$\bar{\mbox{\bf 3}}$ pair, overall color conservation
will guarantee the color singlet value of the other pair. This leads
to a gap formation rate which is similar in lepto- and
hadroproduction, and can be reconciled with the data by introducing a
survival probability. The survival probability accounts for the fact
that gaps can be filled by the underlying event, e.g. mini-jet
production\cite{chehime}, or by higher order processes.  The survival
probability of the gap is expected to be smaller for hadroproduction
than electroproduction, thus accommodating the data.

Discussions of rapidity gap physics have routinely ignored that,
besides quark-quark, gluon-gluon and gluon-quark subprocesses
contribute to jet production in hadron collisions. In the gluon-gluon
color flow diagram corresponding to Fig.~\ref{2j:pom}b top and bottom
protons each split into a color octet gluon and a 3-quark remnant in a
color octet state. There are now $(8 \times 8)^2$ color final states.
Despite the fact that we can at best guess the non-perturbative
dynamics, it is clear that the soft color formalism predicts a gap
rate which is smaller in gluon-gluon interactions. This is in contrast
with the 2-gluon exchange diagram of Fig.~\ref{2j:pom}a which predicts
a gap-rate enhanced by a factor $\left(9 \over 4\right)^2$ in
gluon-gluon subprocesses\cite{zeppo}.  The process is clearly enhanced
when replacing the interacting quarks by gluons, because of the larger
gluon-gluon color coupling.  The contrasting predictions can be easily
tested by enhancing the relative importance of quark-quark
subprocesses in the experimental sample. This can be achieved by
increasing the $p_T$ of the jets at fixed energy, or by decreasing the
collision energy of the hadrons at fixed $p_T$. In either case, we
anticipate an increased rate for the production of gaps in the soft
color scheme; a prediction opposite to that of the 2-gluon exchange
model.  We next confront the contrasting predictions with recent
data\cite{jill}.

Introducing the quantities $F_{QQ}$, $F_{QG}$, and $F_{GG}$ which
represent the frequencies for producing rapidity gaps between a pair
of high-$p_T$ jets in hadronic quark-quark, quark-gluon, and
gluon-gluon collisions, we can write the observed gap fraction as
\begin{equation}
F_{\text gap}(E_T) =
\frac{1}{ d\sigma/dE_T}\left(
  F_{QQ}\frac{d\sigma_{QQ}}{dE_T}
+ F_{QG}\frac{d\sigma_{QG}}{dE_T}
+ F_{GG}\frac{d\sigma_{GG}}{dE_T}\right) \;\; ,
     \end{equation}
where
\begin{equation}
        d\sigma = d\sigma_{QQ} + d\sigma_{QG} + d\sigma_{GG}
\end{equation}
represents the decomposition of the cross section for producing
large-$E_T$ jets into quark-quark, quark-gluon and gluon-gluon
subprocesses.  Predictions can be summarized in terms of the gap
fractions $F_{ij}$. For the 2-gluon hard Pomeron model\cite{zeppo}
\begin{equation}
        F_{QQ}:F_{QG}:F_{GG} = 1:\frac{9}{4}:\left(\frac{9}{4}\right)^2
\;\; .
\label{eq:2-gluon}
\end{equation}
In the soft color calculational scheme the
$F_{ij}$ are independent of the center--of--mass energy and $E_T$,
satisfying
\begin{equation}
        F_{QQ}:F_{QG}:F_{GG} = a:b:c \;\; ,
\end{equation}
with $c < b < a$, in contrast with Eq.~(\ref{eq:2-gluon}). Reasonable
guesses fall in the range
\begin{eqnarray}
        (1/9)^2  < & a~ & < 1/9 \;\; ,
\label{c1} \\
        ac       < & b~ & < \sqrt{ac} \;\; , \\
        (1/64)^2 < & c~ & < 1/64 \;\; .
\label{c3}
\end{eqnarray}

\begin{figure}
\begin{center}
\mbox{\epsfig{file=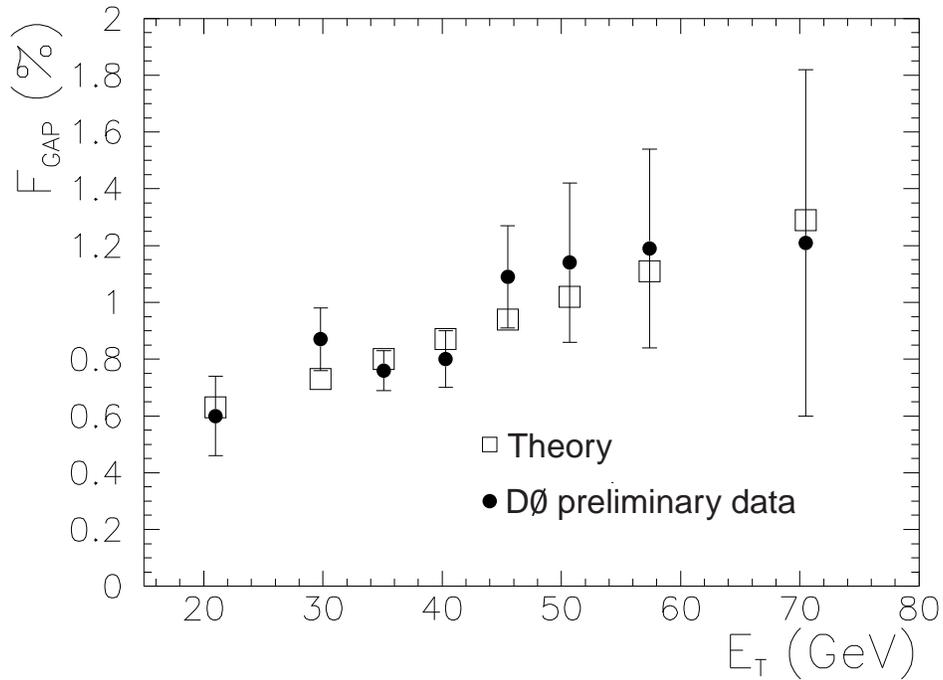,%
              width=0.75\textwidth,%
              angle=0}}
\end{center}
\caption{Theoretical fit to the D\O\ preliminary  data\protect\cite{jill}.
  Notice that each point shown has a different $E_T$ bin, with the
  data being plotted at the mean value of $E_T$ for each bin, which is
  given in Table~\protect\ref{fit}.  }
\label{fig:fit}
\end{figure}

In order to determine the gap fractions $F_{ij}$, we computed the
transverse cross sections $d\sigma_{ij}/dE_T$ in lowest order
perturbative QCD, and subsequently fitted the preliminary D\O\ data
\cite{jill} shown in Fig.~\ref{fig:fit}. We integrated over the $E_T$
bins given in Table~\ref{fit}, imposing that both jets have
$|\eta|>1.9$. The 90\% CL bounds on the $F_{ij}$ are
\begin{eqnarray}
F_{QQ} &=& 0.023^{+ 0.010}
                _{- 0.011}
\;\; ,
 \nonumber \\
F_{QG} &=& 0.00017^{+ 0.012}
                  _{- 0.000}  \;\; ,
\label{eq:param}
\\
F_{GG} &=& 0.0075^{+ 0.0081}
                 _{- 0.0073}
\;\; . \nonumber 
\end{eqnarray}
We also show in Fig.\ \ref{fig:fit} the result of our fit.

The data shows that rapidity gaps are mostly formed in quark-quark
collisions with the value of $F_{QQ}$ exceeding those of $F_{GG}$ and
$F_{GQ}$. The 90\% CL upper limits on $F_{QG}$ and $F_{GG}$ are
$0.012$ and $0.016$, respectively. The dominance of $F_{QQ}$ simple
reflects the experimental fact that the fraction of events with gaps
increases with $E_T$ as the leading order QCD result for
$(d\sigma_{QQ}/dE_T)/(d\sigma/dE_T)$; see Table~\ref{fit}.  Clearly,
the data is consistent with the soft color model for the formation of
rapidity gaps and does not support the 2-gluon exchange approximation
which predicts that processes involving gluons should exhibit larger
rapidity gap frequencies. The data can be interpreted in terms of the
simple color counting previously introduced, which predicts $F_{QQ}
\simeq$ (0.2--0.4)/9 $\simeq$ 2--4 $10^{-2}$ and $F_{GG} \simeq$
(0.2--0.4)/64 $\simeq$ 3--6 $10^{-3}$ for a survival
probability\cite{bj} of 0.2--0.4. This is certainly compatible with
our results given the uncertain systematics of our procedure.

\begin{figure}
\begin{center}
\mbox{\epsfig{file=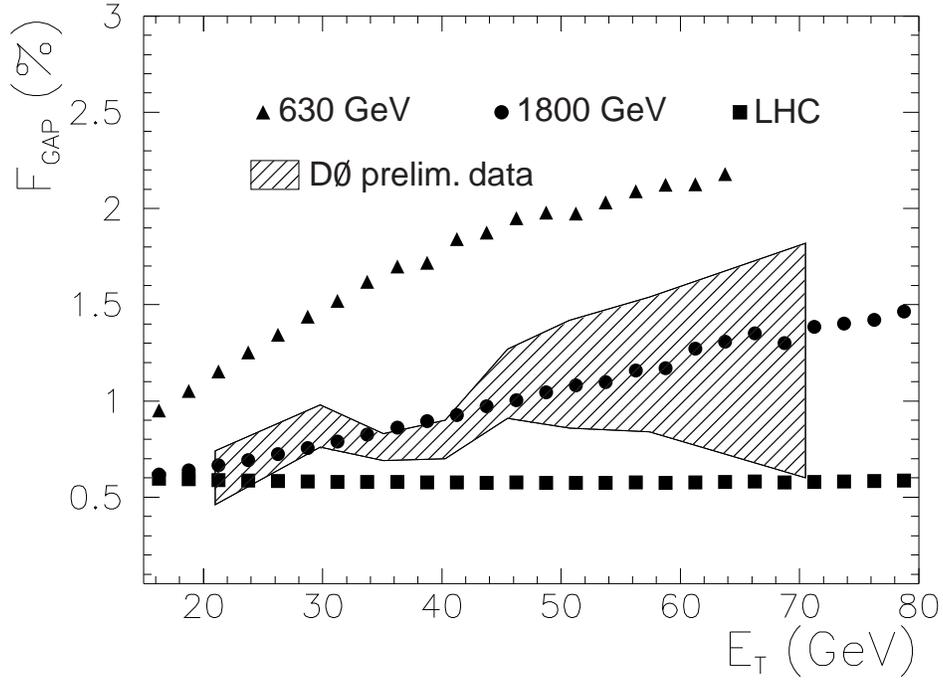,%
              width=0.75\textwidth,%
              angle=0}}
\end{center}
\caption{Our model predictions for rapidity gaps between jets at
Tevatron and LHC.}
\label{fig:bin}
\end{figure}

\begin{figure}
\begin{center}
\mbox{\epsfig{file=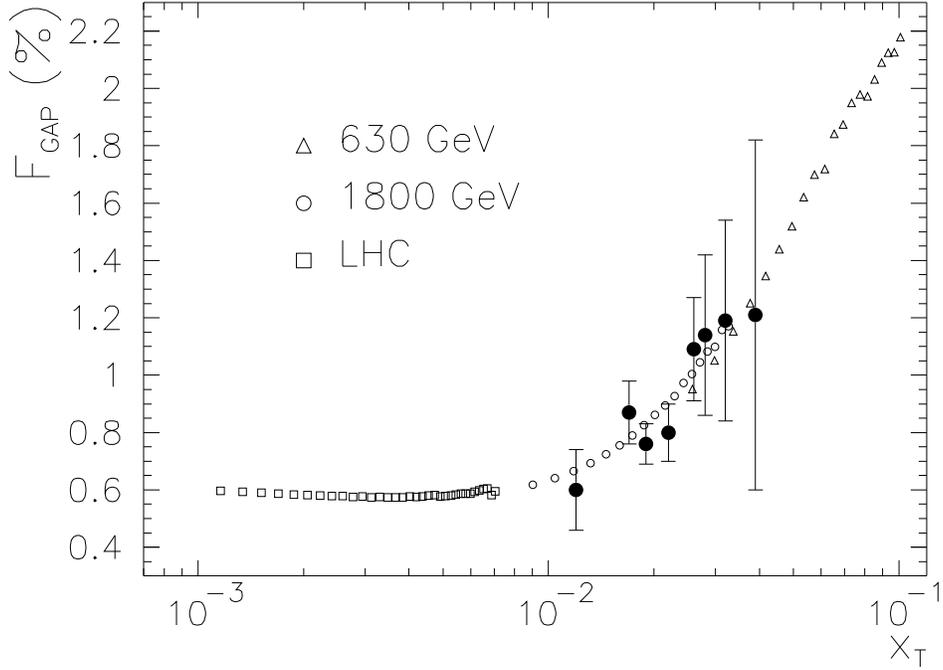,%
              width=0.75\textwidth,%
              angle=0}}
\end{center}
\caption{Same results of Fig.\ \protect\ref{fig:bin} as
  function of the adimensional variable $x_T = E_T/\protect\sqrt{s}$.
  For comparison we exhibit the recent D\O\ data\protect\cite{jill}.
  }
\label{fig:scale}
\end{figure}

Our formalism has non-trivial dynamics built in. It assumes that the
gap fractions $F_{ij}$ are independent of any kinematic variables.
Therefore, the observed fraction of rapidity gaps depends on
kinematical variables only through the relative contributions of the 3
subprocess cross sections $\sigma_{QQ}$, $\sigma_{QG}$, and
$\sigma_{GG}$. The observed gap fractions are, for instance, a
function of $E_T$ only at fixed center--of--mass energy, as shown in
Fig.~\ref{fig:bin}. The $E_T$ distribution was computed using uniform
binning at Tevatron and LHC energies. Furthermore, we also predict
that the gap frequencies are only a function of the scaled variable
$x_T=E_T/\sqrt{s}$. Therefore, the gap frequencies are described by a
universal curve at all energies; see Fig.\ \ref{fig:scale}.

At LHC energy, the fraction of events exhibiting rapidity gaps
associated with moderate $E_T$ jets ($< 80$ GeV) is quite small ($\sim
0.6$\%) due to the large gluon-gluon luminosity.  Notwithstanding, we
anticipate 40 spectacular events for $E_T > 200$ GeV because the gap
frequency increases to $\sim 0.8$\% in this $E_T$ range. On a more
practical note, overlapping events may make such observations
challenging.

\begin{figure}
\begin{center}
\mbox{\epsfig{file=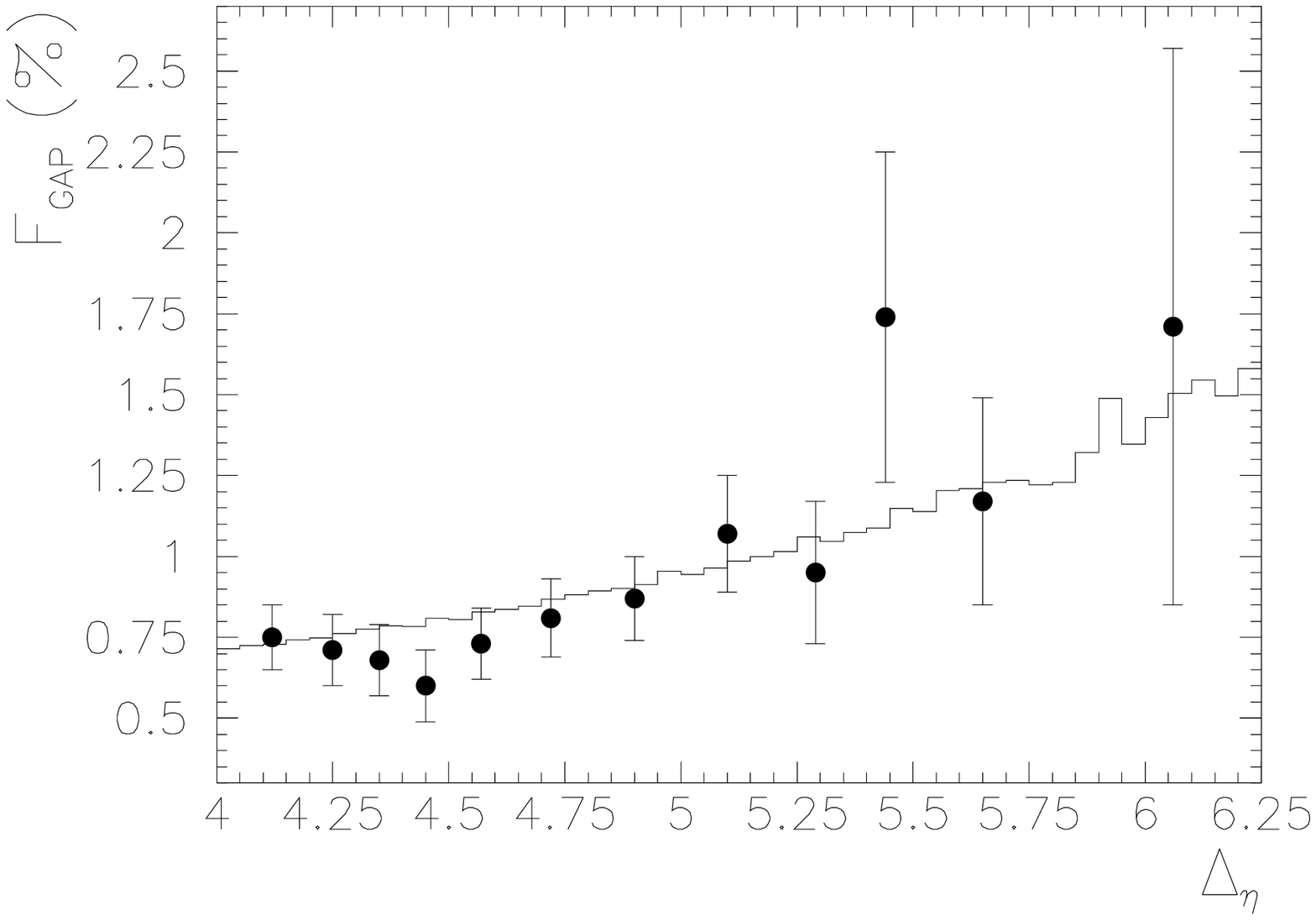,%
              width=0.75\textwidth,%
              angle=0}}
\end{center}
\caption{Comparison with the preliminary D\O\ data\protect\cite{jill} of our
  model prediction for the rapidity gap fraction as function of the
  size of the gap.}
\label{fig:deta}
\end{figure}

The D\O\ Collaboration has also measured the dependence of the
fraction of events with rapidity gaps as a function of the rapidity
separation of the leading jets for $E_T>30$ GeV and $|\eta|>1.7$
\cite{D0:eta}.  Our predictions are successfully confronted with their
recent data\cite{jill} in Fig.~\ref{fig:deta}.

The soft color mechanism can also give rise to rapidity gaps with a
different ordering in rapidity: near the beam are the remnants of an
initial hadron, followed by a rapidity gap and two hard jets which are
separated from the other hadron debris by yet another rapidity gap.
These events represent diffractive gaps. In the context of the color
evaporation model, quark-quark and quark-gluon collisions cannot
initiate such events because of the impossibility to neutralize the
color of the ${\bf 3}$ ($\bar{\bf 3}$) remnants with the exchange of
soft gluons. Therefore, such events only originate in gluon-gluon
collisions, and their fraction should decrease with increasing $E_T$,
or decreasing center-of-mass energy.

We close with some cautionary comments. Do
Figs.~\ref{gap:dis}--\ref{2j:pom} suggest that we have formulated
alternative $s$- and $t$-channel pictures to view the same physics?
Although they seem at first radically different, this may not be the
case. Computation of the exchange of a pair of colorless gluons in the
$t$-channel is not straightforward and embodies all the unsolved
mysteries of constructing the ``Pomeron'' in QCD. In a class of models
where the Pomeron is constructed out of gluons with a dynamically
generated mass\cite{chehime,natale}, the diagram of Fig.~\ref{2j:pom}a
is, not surprisingly, dominated by the configuration where one gluon
is hard and the other soft. The diagram is identical to the standard
perturbative diagram except for the presence of a soft,
long-wavelength gluon whose only role is to bleach color.  Its
dynamical role is minimal, events with gaps are not really different
from events without, suggesting dynamics similar to color
evaporation. Some have argued that in this class of models the hard
Pomeron is no more than an order $\alpha_s^2$ correction, a view which
can be defended on more solid theoretical ground\cite{cudell}. Others
have however challenged the theoretical soundness of this line of
thinking\cite{bj,white}. Also note that our discussion is at best
indirectly relevant to completely non-perturbative phenomena like
elastic scattering.  There is no short distance limit defined by a
large scale. The Pomeron exists.


\acknowledgments

We would like to thank J.~Amundson for collaborations and J.~Bjorken,
M.Drees, G.~Ingelman, A.~White, D~Zeppenfeld, S.~Fleming, B.~May,
J.~Perkins, T.~Taylor, and A.~Brandt for their insight.  This research
was supported in part by the University of Wisconsin Research
Committee with funds granted by the Wisconsin Alumni Research
Foundation, by the U.S.\ Department of Energy under grant
DE-FG02-95ER40896, by Conselho Nacional de Desenvolvimento
Cient\'{\i}fico e Tecnol\'ogico (CNPq), and by Funda\c{c}\~ao de
Amparo \`a Pesquisa do Estado de S\~ao Paulo (FAPESP).


\newpage



\begin{table}[htbp]
\caption{$E_T$ bins used in Fig.~\protect\ref{fig:fit} and its
respective mean value $\bar{E}_T$. For comparison, we also provide the
cross section integrated over the bins for the quark-quark,
quark-gluon, and gluon-gluon subprocesses.}
\begin{center}
\begin{tabular}{|c|c|c|c|c|}
$\bar E_T$ (GeV)& $E_T$ bin (GeV)
&$\int\frac{d\sigma_{QQ}}{dE_T} (pb)$
&$\int\frac{d\sigma_{QG}}{dE_T} (pb)$
&$\int\frac{d\sigma_{GG}}{dE_T} (pb)$
\\ \hline
21.0 & 15-25 & 60.8 & 197 & 163
\\
29.8 & 25-30 & 3.25 & 7.19 & 4.02
\\
35.1 & 30-35 & 1.12 & 2.07 & 0.97
\\
40.3 & 35-40 & 0.43 & 0.68 & 0.27
\\
45.5 & 40-45 & 0.18 & 0.26 & 0.084
\\
50.7 & 45-50 & 0.079 & 0.093 & 0.028
\\
57.4 & 50-60 & 0.056 & 0.057 & 0.014
\\
70.5 & $>60$ & 0.027 & 0.020 & 0.004
\end{tabular}
\label{fit}
\end{center}
\end{table}

\newpage


\begin{references}
\frenchspacing

\bibitem{jill} J.~Perkins (D\O\ Collaboration), {\em Proceedings of
    the 5th International Workshop on Deep Inelastic Scattering and
    QCD}, Chicago, Illinois, 1997; FERMILAB-Conf-97/250-E.
  
\bibitem{faro} O.~\'Eboli, E.~Gregores, and F.~Halzen, {\it
    Proceedings of the 26th International Symposium on Multiparticle
    Dynamics (ISMD\,96)}, Faro, Portugal, 1996 (hep-ph/9611258).

\bibitem{bh}
W.~Buchm\"uller, Phys. Lett. {\bf B353}, 335 (1995);
W.~Buchm\"uller and A.~Hebecker, Phys. Lett. {\bf B355}, 573 (1995).

\bibitem{review}
See, e.g., E.~Braaten, S.~Fleming, and T.C.~Yuan, preprint
OHSTPY-HEP-T-96-001 (hep-ph/9602374), to appear in Ann. Rev. Nucl.
Part. Sci., and references therein.

\bibitem{bbl}
G.T.~Bodwin, E.~Braaten, and G.~Lepage, Phys. Rev. {\bf D51}, 1125 (1995).

\bibitem{cem}
H.~Fritzsch, Phys. Lett. {\bf B67}, 217 (1977).

\bibitem{fh:1a}
F.~Halzen, Phys. Lett. {\bf B69}, 105 (1977).

\bibitem{fh:1b}
F.~Halzen and S.~Matsuda, Phys. Rev. {\bf D17}, 1344 (1978).

\bibitem{gor}
M.~Gluck, J.~Owens, and E.~Reya, Phys. Rev. {\bf D17}, 2324 (1978).

\bibitem{amundson}
J.~Amundson, O.~\'Eboli, E.~Gregores, and F.~Halzen, Phys.Lett. {\bf
B372}, 127 (1996).

\bibitem{oscar}
J.~Amundson, O.~\'Eboli, E.~Gregores, and F.~Halzen, Phys. Lett.
{\bf B390}, 323 (1997).

\bibitem{oscarZ}
O.~\'Eboli, E.~Gregores, and F.~Halzen, Phys. Lett. {\bf B395}, 113
(1997).

\bibitem{schuler}
G.A.~Schuler, preprint CERN-TH.7170/94 (hep-ph/9403387).

\bibitem{prl1} S.~Abachi {\em et al.} (D\O\ Collaboration),
Phys. Rev. Lett. {\bf 72}, 2332 (1994).

\bibitem{cdf1} F.~Abe {\em et al.} (CDF Collaboration),
Phys. Rev. Lett. {\bf 74}, 855 (1995).

\bibitem{prl2} S.~Abachi {\em et al.} (D\O\ Collaboration),
Phys. Rev. Lett. {\bf 76}, 734 (1996).

\bibitem{chehime}
H.~Chehime, {\em et al.,} Phys. Lett. {\bf B286} 397 (1992).

\bibitem{zeppo}
D. Zeppenfeld, preprint MADPH-95-933 (hep-ph/9603315).

\bibitem{bj}
J.D.~Bjorken, Int. J. Mod. Phys. {\bf A7} 4189 (1992); Phys.
Rev. {\bf D47} 101 (1993); 
preprint SLAC-PUB-5823 (1992).

\bibitem{D0:eta}
T.\ Thomas, talk given at the {\em 3rd Workshop on Small x and
Diffractive Physics}, Argonne, Sept.~26--29, 1996.

\bibitem{natale}
A.A.~Natale, {\em et al.} Phys. Rev. {\bf D47} (1993) 295; {\bf
D48} 2324 (1993).

\bibitem{cudell}
J.R.~Cudell, A.~Donnaichie, and P.V.~Landshoff, to appear.

\bibitem{white}
A.~White, talk given at the {\em 3rd Workshop on Small x and
Diffractive Physics}, Argonne, Sept.~26--29, 1996.

\end{references}
\end{document}